# Rabi-Bloch Oscillations in Spatially Distributed Systems: Temporal Dynamics and Frequency Spectra

Ilay Levie, Rafael Kastner and Gregory Slepyan[*]

*School of Electrical Engineering, Tel Aviv University, Tel Aviv, Israel*

We considered one-dimensional chain of the two-level quantum systems coupled via tunneling. The chain is driven by the superposition of dc and ac fields in the strong coupling regime. Based on the fundamental principles of electrodynamics and quantum theory, we developed a generalized model of quantum dynamics for such interactions, free of rotating wave approximation (RWA). The system of motion equations was studied numerically. We analyzed the dynamics and spectra of inversion density, dipole current density and tunneling current density. In the case of resonant interaction with ac-component the particle dynamics exhibits itself in the oscillatory regime, which may be interpreted as a combination of Rabi- and Bloch oscillations with their strong mutual influence. Such scenario for an obliquely incident ac field dramatically differs from the individual picture both types of oscillations due to an interactions. This novel effect is counterintuitive because of the strongly different frequency ranges for such two types of oscillations existence. This dynamics manifests itself in multi-line spectra at different combinations of Rabi- and Bloch frequencies. The effect is promising as a framework of new type of spectroscopy in nanoelectronics and electrical control of nano-devices.

## I. INTRODUCTION

Periodic low-dimensional lattices been a platform for many quantum phenomena since appearance in the famous models of ferromagnetism in statistical mechanics (Ising chain [1], Heisenberg spin-chain [2]) and the simplest model of interacting particles in a lattice to describe the transition between conducting and insulating systems (Habbard chain [3]). Almost centenary history its further development went on the ways to address the various forms of periodicity (a lattice in position space, a lattice in momentum space, time-periodic forces - lattice in time, and their different combinations), the various origin of lattice element (atom, ion, molecule, quantum dot) and mechanism of their chain-formed interaction (dipole-dipole interactions, exchange interactions, Coulomb interactions, tunneling, etc.). The great progress in atomic optics and nano-technologies made experimentally feasible a great number of manifold physical systems of a given type: semiconducting and graphene superlattices [4,5], Bose-Einstein condensate trapped in an optical lattice [6], trapped ions chain [7,8], cooled atomic array in the cavity [9], the array of Josephson contacts [10], the crystalline arrays of quantum dots connected by conducting chains of linker atoms or interacted via another physical mechanisms [11-15], conjugated polymer chains [16], a lattice of p-wave superconductor in 1D (Kitaev chain) [17], etc. All the systems mentioned above are characterized by a number of qualitative features that define the properties of their energy spectra and transport properties, which are important from the point of view of interaction with classical and quantum electromagnetic fields in the microwave, terahertz and visible frequency ranges. These properties of low-dimensional nanostructures differentiate them from traditional bulk materials and open the future prospects of their applications in nano-electronics and nano-photonics, quantum informatics and quantum computing [18].

Recent theoretical progress in this area is associated with the novelty of array types and inter-element interactions. Among them one may note the mechanisms of dissipative inter-element interactions [19] and noisy coupling [20], which manifest themselves in surprising thermodynamic behavior of the arrays. Scully with his team [21] developed the concept of the superradiance lattice, which corresponds to a collection of three-level atoms in timed Dicke states constructed based on[1] electromagnetically induced transparency. In particular, the superradiance lattice can be extended to three and higher dimensions, with new physics waiting to be explored. Strong coupling of light with spatially distributed chains of two-level fermionic systems manifests itself in the spatial movement of Rabi-oscillations (RO) in the form of the quantum transition wave (Rabi-wave), where the light plays a role of refractive medium for its propagation [22-27]. Among the promising applications of Rabi-waves we note the electrically tunable optical nano-antennas with highly directive emission [22,24,27].

The dynamics of a single particle in a periodic potential dramatically changes under the influence of a static force. The question how electrons would behave in a crystal lattice once a dc electric field is applied was raised by Bloch [30] and Zener [31]. When electrons in crystalline potentials are subject to uniform external fields, Bloch predicted that the quantum coherence properties of the electrons would prevent their transport. He showed that the electrons dynamically localize and undergo periodic oscillations in space (Bloch oscillations, BO). The effect is non-intuitive: one observes a periodic dynamics although the periodic structure is violated instead of the expected acceleration towards infinity due to the dc-field. It leads to the existence of localized modes (Wannier–Stark states) with equidistant wave-number spacing (Wannier–Stark ladder) that do not undergo diffraction [32,33]. The BO frequency is $\Omega_B = eaE_{Dc}$, where $e$ is electron charge, $a$ is the lattice period and $E_{Dc}$ is the dc-field value [24,35].

Remarkably, BO manifests the wave properties of the electrons, and therefore appears in other types of waves in tilted periodic potentials. Inspired by the technological achievements in semiconductor growth at the end of 20-th century, similar phenomena were found in other physical systems, e.g., for electronic wave packets in semiconductor supperlattices [36,37], cold atoms in optical lattices [38,39], light beams in periodic structures [40,41] and waveguide arrays with linearly varying propagation constants [42, 43] (optical BO). It was shown [44] that BO are able to manifest itself in quantum optics too: when photons in N-particle path-entangled undergo Bloch oscillations, they exhibit a periodic transition between spatially bunched and antibunched states. The period of the bunching-antibunching oscillation is N times faster than the period of the photon density oscillation, manifesting their unique coherence properties. Recent progress in fabrication of the waveguide lattices, photon-number resolving detectors and photonic entangled state sources [45, 46], made its experimental observation within reach [47].

Despite their apparent simplicity, the dynamics of quantum particles in periodic structures becomes full of surprises due to the synthesis of different physical mechanisms. One of such examples is the array driven by dc and ac linear (in $x$) electric fields simultaneously: $V(x,t) = -x(E_0 + E_1 \cos(\omega t + \varphi))$, where $V(x,t)$ is the total applied potential, $E_0, E_1$ are given constant amplitudes, $\varphi$ is a constant phase. A resonant condition $\omega \approx \Omega_B$ results in a beat between the Bloch cycle and the drive, with a drastic change in the particle motion. On top of the Bloch oscillation, the intraband motion shows a much larger oscillation in position space that extends over hundreds of lattice sites [48]. Those "super-Bloch oscillations" directly correspond to the motion of normal Bloch oscillations, just rescaled in space and time. Several recent studies [47,49-54] have investigated the dynamics of cold atoms in optical lattices subject to potential ac-field forcing; the theoretically predicted renormalization of the tunneling current amplitudes [55] has been verified experimentally.

The subject of this paper is the synthesis of BO and RO. We consider a lattice of two-level fermionic systems which is driven by dc and ac electric fields. A resonant condition is $\omega \approx \omega_0$, where $\omega_0$ is the inter-level transition frequency. In contrast with super-Bloch oscillations, the motion here is the combined dynamics of intra-band and inter-band transitions called the Rabi-Bloch oscillations (RBO). The total current is the sum of low-frequency component (tunneling) and high-frequency component (optical transitions). The total field at the chain axis is

*gregory_slepyan@yahoo.com

$\mathbf{E}(x,t) = -\mathbf{e}(\partial V_{Dc}/\partial x) + \mathbf{E_0}\cos(\omega t - kx + \varphi)$, where $\mathbf{e}$ is a unit vector along the atomic chain, $k$ is a component of the wave vector along the chain axis, $\mathbf{E_0}$ is the ac-field amplitude and $V_{Dc}$ is the static potential. The ac field is non-gradient – it represents the wave, travelling over the chain. This wave propagation results in a combined effective lattice, produced both in the position space and in time. This lattice manifests itself in strong and manifold interaction of Rabi and Bloch oscillations in spite of their strongly different frequency ranges. Such interactions are considered in detail below and represent the main result of our paper. A simplified RBO scenario was considered in [56], however with ac field assumed to be homogeneous along the chain ($k=0$), which corresponds to the effective lattice in time only and stops short the main part of effects.

It is well known, that a coherent superposition of atomic states in three-level atom in $\Lambda$-configuration is responsible for such interesting phenomenon as a coherent trapping [57]. These atoms are effectively transparent to the incident field and have such important applications as lasing without inversion, refractive index enhancement in a non-absorbing medium and electromagnetically induced transparency [57]. As it will be shown in this paper, the similar states exist in the chains of two-level atoms. The role of destructive quantum interference between two lower states in this case plays inter-atomic tunneling. It will be shown up, that RBO for such type of states differ in a number of features.

Our paper is organized in the following way. In Sec. II, the model system is introduced and equations of its dynamical behavior are derived. In Sec. III, the numerical results for inversion, tunneling current and displacement current are presented and discussed from physical point of view. In Sec. IV validity of the RWA approximation has been considered. In Sec. V the RBO scenario for the coherently trapped states has been investigated. Finally, the main results and outlook problems are formulated in Sec.VI.

## II. MODEL

Let us consider one-dimensional (1D) structure of identical atoms uniformly distributed over linear lattice points with period $a$, see Fig.(1). Each atom is considered as a two-level Fermion system with transition frequency $\omega_0$. For brevity, we refer to the arbitrary two-level quantum object as an "atom" regardless of its physical nature, e.g., quantum dot (QD), polar molecule, trapped ion, etc. The location of the atoms in the lattice points is determined by the radius-vector $\mathbf{R}_j = \mathbf{e}ja$, $j=0,1,\ldots,N$, $N$ is a number of atoms. The index $j$ completely determines the atom location. We assume the atoms to be coupled via the inter-atomic tunneling. The chain is driven by electrostatic (dc) field directed along the axis and simultaneously interacts with monochromatic electromagnetic (ac) field. We will consider the case of dipole interaction in the regime of strong coupling and assume the resonant condition to be fulfilled ($\omega_0 \approx \omega$). The system under consideration exhibits complex single-particle RBO, for which the theoretical framework is introduced below.

### A. Unperturbed Hamiltonian in Wannier basis

Theoretical analysis is based on the single-particle Hamilton approach with the Hamiltonian $\hat{H} = \hat{H}_0 + \hat{H}_I$, where

$$\hat{H}_0 = \frac{\hat{\mathbf{p}}^2}{2m} + V(\mathbf{r}) \tag{1}$$

is the component of free electron movement in the chain associated with the interatomic tunneling, which is described by the periodic potential $V(\mathbf{r})$. The excited and ground states of the atoms here and thereafter will be denoted *a* and *b*, respectively. The eigenmodes of Hamiltonian (1) are two Bloch modes correspondent to valence (*b*) and conductive (*a*) zones and denoted as $\psi_{n,h}(\mathbf{r}) = |n,h\rangle$, $n=a,b$, $h$ being a scalar quasi-momentum directed along the chain. The second component of the Hamiltonian

$$\hat{H}_I = -e\mathbf{E}(\mathbf{r},t) \cdot \hat{\mathbf{r}} \qquad (2)$$

describes the dipole interaction of the chain with the total (dc and ac) electromagnetic field, $\hat{\mathbf{r}}$ is the operator of electron position in the chain.

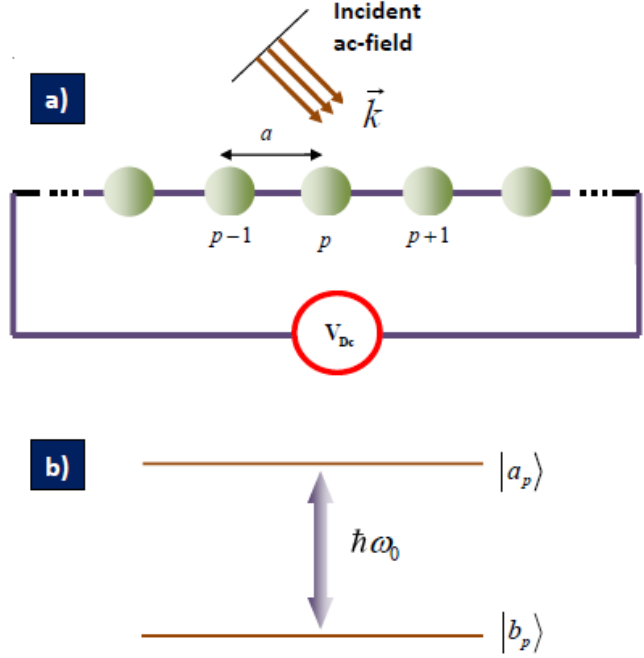

FIG. 1. a) General illustration of the periodic two-level atomic chain used as a model, indicating Rabi-Bloch oscillations. It is excited with obliquely incident ac field in the strong coupling regime and driven with dc voltage applied at the ends. The neighboring atoms are coupled via interatomic tunneling with different values of penetration at the ground and excited states. b) Ground and excited energy levels of single two-level atom separated by the transition energy $\hbar\omega_0$.

Because the electrons are strongly confined inside the atoms, we will use as a basis Wannier functions $\phi_{n,\mathbf{R}_j}(\mathbf{r}) = |n,\mathbf{R}_j\rangle$ [59], defined as a linear combinations of Bloch modes

$$\phi_{n,\mathbf{R}_j}(\mathbf{r}) = \frac{1}{\sqrt{N}} \sum_h e^{-ih(\mathbf{e}\cdot\mathbf{R}_j)} \psi_{n,h}(\mathbf{r}) \qquad (3)$$

Let us mention some properties of Wannier functions, which are important for our analysis. Wannier function $\phi_{n,\mathbf{R}_j}(\mathbf{r})$ is strongly localized inside the *j*-th atom. Inversion of eq (3) gives the expression of Bloch states in the terms of Wannier functions:

$$\psi_{n,h}(\mathbf{r}) = \frac{1}{\sqrt{N}} \sum_{\mathbf{R}_j} e^{ih(\mathbf{e}\cdot\mathbf{R}_j)} \phi_{n,\mathbf{R}_j}(\mathbf{r} - \mathbf{R}_j) \qquad (4)$$

Wannier functions at different locations are related via relationship $\phi_{n,\mathbf{R}_j}(\mathbf{r}) = \phi_{n,\mathbf{R}_j+\mathbf{R}_l}(\mathbf{r}+\mathbf{R}_l)$ and satisfy the orthogonality relation $\langle n, \mathbf{R}_j | m, \mathbf{R}_l \rangle = \delta_{mn}\delta_{\mathbf{R}_j,\mathbf{R}_l}$.

The first step of our analysis is the presentation of Hamiltonian (1) in the Wannier basis. Projecting $\hat{H}_0$ on the Bloch modes and using (4), we obtain the dispersion law for free-tunneling electron in the next form:

$$\varepsilon_n(h) = \langle n, h | \hat{H}_0 | n, h \rangle = \frac{1}{N} \sum_{\mathbf{R}_j, \mathbf{R}_l} e^{-ih\mathbf{e}\cdot(\mathbf{R}_j-\mathbf{R}_l)} \langle n, \mathbf{R}_j | \hat{H}_0 | n, \mathbf{R}_l \rangle \tag{5}$$

Denoting

$$\langle n, \mathbf{R}_j | \hat{H}_0 | n, \mathbf{R}_l \rangle = \begin{cases} \varepsilon_n(0), & j = l \\ t_n(\mathbf{R}_j - \mathbf{R}_l), & j \neq l \end{cases} \tag{6}$$

we can rewrite (5) in the following form:

$$\varepsilon_n(h) = \varepsilon_n(0) + \frac{1}{N} \sum_{\mathbf{R}_j, \mathbf{R}_l, j \neq l} e^{-ih\mathbf{e}\cdot(\mathbf{R}_j-\mathbf{R}_l)} t_n(\mathbf{R}_j - \mathbf{R}_l) \tag{7}$$

where $t_{a,b}(\mathbf{R}_j - \mathbf{R}_l) = t^*_{a,b}(\mathbf{R}_l - \mathbf{R}_j)$. The Hamiltonian (1) may be presented with the Wannier basis in the form of a block diagonal matrix

$$\hat{H}_0 = \begin{pmatrix} \hat{H}_{0a} & 0 \\ 0 & \hat{H}_{0b} \end{pmatrix} \tag{8}$$

with diagonal-on sub-matrixes

$$\hat{H}_{0n} = \varepsilon_n(0) \sum_{\mathbf{R}_j} |n, \mathbf{R}_j\rangle\langle n, \mathbf{R}_j| + \sum_{\mathbf{R}_j, \mathbf{R}_l, j \neq l} t_n(\mathbf{R}_j - \mathbf{R}_l) |n, \mathbf{R}_l\rangle\langle n, \mathbf{R}_j| \tag{9}$$

We assume that tunneling coupling exist only between the neighboring atoms (tight-binding approximation). Thus, we omit for brevity the spatial arguments in tunneling matrix elements: $t_{a,b}(\mathbf{R}_j - \mathbf{R}_{j+1}) = t_{a,b}$, $t_{a,b}(\mathbf{R}_j - \mathbf{R}_{j-1}) = t^*_{a,b}$. The Hamiltonian (8) subject to (9) reads

$$\hat{H}_0 = \varepsilon_b(0) \sum_j |b, j\rangle\langle b, j| + \varepsilon_a(0) \sum_j |a, j\rangle\langle a, j| +$$
$$\sum_j (t_b |b, j\rangle\langle b, j+1| + t^*_b |b, j\rangle\langle b, j-1|) + \tag{10}$$
$$\sum_j (t_a |a, j\rangle\langle a, j+1| + t^*_a |a, j\rangle\langle a, j-1|)$$

where Wannier functions re-denoted in conformity with $|n, \mathbf{R}_j\rangle \to |n, j\rangle$.

### B. Interaction Hamiltonian in Wannier basis

Let us proceed to the calculation of matrix elements $\langle m,l|\hat{H}_I|n,j\rangle$ for the interaction Hamiltonian (2). Such a motion as a combined interband transition and interatomic tunneling through the potential barrier requires so high values of energy, that its probability becomes negligibly small. Thus, for the simultaneously obeyed inequalities $m \neq n, l \neq j$ we have $\langle m,l|\hat{H}_I|n,j\rangle \approx 0$. We assume the intra-atomic confinement sufficiently high, such that electric field is approximately homogeneous at the support area of the localized Wannier function. Thus,

$$\langle m,j|\hat{H}_I|n,j\rangle \approx -e\mathbf{E}(ja,t)\cdot\langle m,j|\hat{\mathbf{r}}_j|n,j\rangle \qquad (11)$$

with the electric field

$$\mathbf{E}(j,t) \approx \mathbf{E}_{Dc} + \omega A_0 \mathbf{u}\cos[(\mathbf{k}\cdot\mathbf{e})ja - \omega t], \qquad (12)$$

where the first term is a bias dc field, the second one is the incident plane wave, $\mathbf{E}_{Dc}$ and $A_0$ are constant values, $\mathbf{u}$ is the unit vector of polarization, $\mathbf{k}$ is the wave vector. The plane wave in general case is oblique with respect to the chain axis ($\mathbf{k}\cdot\mathbf{e} \neq 0$), thus the electron moves in a spatial-temporal lattice produced by the second term in Eq. (12). Let us shift the origin for a given $j$ to the center of the $j$-th atom via the relation $\hat{\mathbf{r}}_j = \mathbf{R}_j \hat{I} + \hat{\mathbf{r}}'$. Using the orthogonality of the Wannier functions for different bands, we have $\langle m,j|\hat{\mathbf{r}}_j|n,j\rangle = \mathbf{R}_j \delta_{mn} + \langle m,j|\hat{\mathbf{r}}'|n,j\rangle$, which makes (11)

$$\langle m,j|\hat{H}_I|n,j\rangle \approx -\mathbf{E}(ja,t)\cdot\left(e\mathbf{R}_j\delta_{mn} + \mathbf{d}_{mn}\right) \qquad (13)$$

where

$$\mathbf{d}_{mn} = e\langle m,j|\hat{\mathbf{r}}'|n,j\rangle \qquad (14)$$

It is important to note that the coefficients (14) are position-independent. Another type of matrix elements required for our analysis couples the electrons of neighboring atoms in the same band via the ac field (tunneling photon assistance):

$$\langle n,j|\hat{H}_I|n,j+1\rangle = -e\omega A_0 d_{nj}(t) \qquad (15)$$

where

$$d_{nj}(t) = \langle n,j|(\hat{\mathbf{r}}\cdot\mathbf{u})\cos((\mathbf{k}\cdot\hat{\mathbf{r}})-\omega t)|n,j+1\rangle \approx \cos((\mathbf{k}\cdot\mathbf{e})ja-\omega t)\langle n,j|(\hat{\mathbf{r}}\cdot\mathbf{u})|n,j+1\rangle \qquad (16)$$

### C. Equations of motion

Our starting point is the Schrodinger equation $i\hbar\partial_t|\psi\rangle = \hat{H}|\psi\rangle$ with the total Hamiltonian being

$$\hat{H} = \varepsilon_b(0)\sum_j |b,j\rangle\langle b,j| + \varepsilon_a(0)\sum_j |a,j\rangle\langle a,j| +$$

$$\sum_j (t_b |b,j\rangle\langle b,j+1| + t_b^* |b,j\rangle\langle b,j-1|) + \sum_j (t_a |a,j\rangle\langle a,j+1| + t_a^* |a,j\rangle\langle a,j-1|) -$$

$$\sum_j \mathbf{E}(j,t)\cdot(e\mathbf{R}_j + \mathbf{d}_{bb})|b,j\rangle\langle b,j| - \sum_j \mathbf{E}(j,t)\cdot(e\mathbf{R}_j + \mathbf{d}_{aa})|a,j\rangle\langle a,j| -$$

$$\omega A_0 \sum_j d_{bj}(t)|b,j\rangle\langle b,j+1| - \omega A_0 \sum_j d_{aj}(t)|a,j\rangle\langle a,j+1| - \quad (17)$$

$$\omega A_0 \sum_j d_{bj}^*(t)|b,j\rangle\langle b,j-1| - \omega A_0 \sum_j d_{aj}^*(t)|a,j\rangle\langle a,j-1| -$$

$$\sum_j \mathbf{E}(j,t)\cdot\mathbf{d}_{ab}|a,j\rangle\langle b,j| - \sum_j \mathbf{E}(j,t)\cdot\mathbf{d}_{ba}|b,j\rangle\langle a,j|,$$

which follows from (10), (13)-(16). We present the required wave function as a superposition of Wannier functions

$$|\psi(t)\rangle = \sum_j \{a_j(t)|a,j\rangle + b_j(t)|b,j\rangle\} \quad (18)$$

with unknown coefficients $a_j(t), b_j(t)$. Using the orthogonality of Wannier functions mentioned above, we obtain the system of coupled differential equations for the probability amplitudes $a_j(t), b_j(t)$:

$$i\hbar\frac{\partial a_j}{\partial t} = \left(\varepsilon_0 + \delta\varepsilon - \mathbf{E}(j,t)\cdot(e\mathbf{R}_j + \mathbf{d}_{aa})\right)a_j +$$
$$\left(t_a - \omega A_0 d_{aj}(t)\right)a_{j+1} + \left(t_a^* - \omega A_0 d_{aj}^*(t)\right)a_{j-1} - \mathbf{E}(j,t)\cdot\mathbf{d}_{ab}b_j \quad (19)$$

$$i\hbar\frac{\partial b_j}{\partial t} = \left(\varepsilon_0 - \delta\varepsilon - \mathbf{E}(j,t)\cdot(e\mathbf{R}_j + \mathbf{d}_{bb})\right)b_j +$$
$$\left(t_b - \omega A_0 d_{bj}(t)\right)b_{j+1} + \left(t_b^* - \omega A_0 d_{bj}^*(t)\right)b_{j-1} - \mathbf{E}(j,t)\cdot\mathbf{d}_{ba}a_j \quad (20)$$

where $\varepsilon_0 = (\varepsilon_a(0) + \varepsilon_b(0))/2$, $\delta\varepsilon = (\varepsilon_a(0) - \varepsilon_b(0))/2$.

The system (19),(20) represents the framework for investigation of Rabi-Bloch oscillations. It is analytically unsolvable; therefore it will be integrated numerically. It is free from some conventional approximations (for example, RWA [57]). The coefficients $t_{a,b}, d_{aa,bb}, d_{aj,bj}, d_{ab}$ will be considered as phenomenological apriori given parameters. It may be instructive to dwell on the physical meaning of the different terms and values appeared in the system (19),(20). The quantity $\delta\varepsilon$ is the minimal energy of the optical transition between the valence and conductive bands. It defines the frequency of free interband oscillations in the absence of EM field. The energy benchmark given by the value $\varepsilon_0$ defines the phase factor, which doesn't supports observable values, and may be set as $\varepsilon_0 = 0$ [60]. The factor $\mathbf{E}(j,t)\cdot\mathbf{R}_j \cong ja\mathbf{E}(j,t)\cdot\mathbf{e}$ in (19) and (20) describes Bloch and super-Bloch oscillations at the excited and ground states, respectively [35,48]. The coefficients $d_{aa,bb}$ vanish for the real atoms (excluding hydrogen atom) on the assumption of symmetry properties [60]. Hovewer, in some cases may be non-zero for such artificial atoms, as semiconductor quantum dots with broken inversion symmetry [61]. It leads to the appearance of an additional line in the spectrum of Rabi-oscillations for the single quantum dot [61] and will be considered below for the case of the particle chain. The factors $t_{a,b} - \omega A_0 d_{aj,bj}(t)$ describe the coupling between the neighboring atoms (the first term corresponds to the direct tunneling; the second one exhibits the tunneling photonic assistance). The last terms in (19),(20) couple the ground and excited states via the EM-field. The value $\Omega_R = (\mathbf{u} \cdot \mathbf{d}_{ab})\omega A_0 / \hbar$ may be associated with the Rabi-frequency for the chain. It differs from the conventional Rabi-frequency defined for separate atom [57], because of the appearance in (14) the Wannier states instead of atomic orbitals. For the deep atomic levels the Rabi-frequency [57] approximates the conventional Rabi-frequency, because the deep Wannier states tend to the correspondent atomic orbitals.

The system (19), (20) may be simplified assuming the atoms to be inversion symmetrical ($\mathbf{d}_{aa} = \mathbf{d}_{bb} = 0$) and neglecting the photon assistance ($d_{aj,bj} \approx 0$). In this case these equations become

$$i\hbar \frac{\partial a_j}{\partial t} = \left(\delta\varepsilon - \mathbf{E}_0 \cdot e\mathbf{R}_j\right)a_j + t_a a_{j+1} + t_a^* a_{j-1} - \mathbf{E}_j(t) \cdot \mathbf{d}_{ab} b_j, \tag{21}$$

$$i\hbar \frac{\partial b_j}{\partial t} = -\left(\delta\varepsilon + \mathbf{E}_0 \cdot e\mathbf{R}_j\right)b_p + t_b b_{j+1} + t_b^* b_{j-1} - \mathbf{E}_j(t) \cdot \mathbf{d}_{ba} a_j \tag{22}$$

### D. Inversion and electric current

The central quantities describing the Rabi-Bloch oscillations are inversion, tunneling and dipole currents, expressed in the terms of probability amplitudes, to be calculated from the eqs (19), (20) (or (21),(22)). These values are position-dependent, therefore we will refer to their densities per unit cell of the chain. The inversion density is

$$w_j(t) = \frac{1}{N}\left(|a_j|^2 - |b_j|^2\right) \tag{23}$$

The equation for tunneling current density is

$$J_{\text{Tunneling},j}(t) = J^{(a)}_{\text{Tunneling},j}(t) + J^{(b)}_{\text{Tunneling},j}(t) = \\ -i\frac{e}{2}t_a\left(a_{j-1}(t) - a_{j+1}(t)\right)a_j^*(t) - i\frac{e}{2}t_b\left(b_{j-1}(t) - b_{j+1}(t)\right)b_j^*(t) + c.c. \tag{24}$$

(Detail derivation of (24) is given in the Appendix).

The operator of dipole current in the Heisenberg picture is $\hat{J}_{\text{Dipole},j}(t) = c\partial \hat{P}_j(t)/\partial t$, where $\hat{P}_j(t)$ is the polarization operator at the point $j$. At the Schrodinger picture we obtain $\hat{J}_{\text{Dipole},j} = -i\omega c e d_{ab}|a,j\rangle\langle b,j| + H.c.$, and the observable current reads

$$J_{\text{Dipole},j}(t) = -i\omega c e d_{ab} a_j^*(t) b_j(t) + c.c. \tag{25}$$

## III. RESULTS AND DISCUSSION

The system under consideration comprises a large number of physical phenomena. For convenience, we define certain combinations of these phenomena, which correspond to different physical regimes presented in Table I.

Let us assume that the system is initialized with excited Gaussian wavepacket

$$a_j(0) = g e^{-\frac{(j-j')^2 a^2}{\sigma^2}} \qquad (26)$$

$$b_j(0) = 0 \qquad (27)$$

where $g$ is normalization factor, $j', \sigma$ are the position of Gaussian center and effective Gaussian width, respectively. Here we will present the results, obtained via the numerical solution of eqs (21), (22).

TABLE I; Classification of possible interaction types and correspondent dynamics in the 1D-chain with RBO

| Variant | The type of interaction | The type of dynamics |
|---|---|---|
| a) | $E_{Dc} \neq 0, E_{Ac} = 0$ | Bloch oscillations |
| b) | $E_{Dc} = 0, E_{Ac} \neq 0, k = 0$ | Rabi-oscillations |
| c) | $E_{Dc} = 0, E_{Ac} \neq 0, k \neq 0$ | Rabi Waves |
| d) | $E_{Dc} \neq 0, E_{Ac} \neq 0, k = 0$ | Rabi-Bloch oscillations in the standing ac-field |
| e) | $E_{Dc} \neq 0, E_{Ac} \neq 0, k \neq 0$, $t_a = t_b$ | Rabi-Bloch oscillations in the travelling wave; both ground and excited states are near the top of quantum box (Rydberg-like type of spectrum) |
| f) | $E_{Dc} \neq 0, E_{Ac} \neq 0, k = 0$ $t_a \gg t_b$ | Rabi-Bloch oscillations in the standing ac-field; ground state energy level near the bottom of quantum box |

Numerical modeling shows that the system dynamics represent the superposition of two phenomena: the first one is the inter-atomic intra-band transitions via electron tunneling; the second one is the intra-atomic inter-band quantum transitions. In spite of this components belonging to different frequency ranges, the total motion does not add up to their simple linear superposition. It is characterized by their strong mutual influence, which produces non-trivial dynamics and qualitative features of spectra. The reason for it is the diffraction by the spatial-temporal lattice induced by the ac field (the factor $\cos\left[(\mathbf{k}\cdot\mathbf{e})ja - \omega t\right]$ in the last terms of (21), (22)).

Fig. 2 shows the temporal behavior of inversion density. Let us start with two illustrative limit cases: for the first one $A_0 = 0$ (no ac field and hence no Rabi oscillations). It corresponds to the ordinary BO: inversion oscillates at the lower Bloch frequency $\Omega_B$ [34] (Fig 2a). The oscillations are approximately monochromatic: the values of high harmonics is rather weak (see Fig 4a). For the second case $\mathbf{E}_0 = 0$ (the driven dc field is absent- Fig 2b). This case corresponds to the ordinary RO [57]; the inversion oscillates monochromatically at the higher Rabi-frequency $\Omega_R$. Fig 2d shows the first example of RBO: both ac and dc fields are present. The ac field normally incident with respect to the chain axis ($k = (\mathbf{k}\cdot\mathbf{e}) = 0$), therefore the ac-field is homogeneous along the axis. This is the case of temporal lattice, in which the total motion adds up to a linear superposition of RO and BO. In other words, RO become modulated by the Bloch frequency. The inversion spectrum is transformed to the triplet (Fig 4d): the central line with $\omega = \Omega_R$ and two side bands with $\omega = \Omega_R \pm \Omega_B$.

The qualitative behavior of motion dramatically changes for spatial-temporal lattice, where the ac field is obliquely incident ($k \neq 0$) – see Figs 2c,e. The initial Gaussian packet rapidly decays over a set of rather small sub-packets, thus the state becomes a strongly oscillating packet with approximately Gaussian envelope. It is manifestation of the diffraction mentioned above, along with the spatial oscillations correspond to the Floquet-harmonics. As a result, the BO is accompanied with sub-packets relative motion with respect to another. This motion leads to appearance of additional spectral lines (Figs 4c,e) with amplitudes comparable to the main one. Fig 2d corresponds, for example, to the case of Rydberg atom – two sidebands appear in excited states with very high principal quantum numbers [62]. As a result, the energies of two states there are placed near the edge of potential box, and the probabilities of their penetration over the potential barrier are approximately equal (for simplicity, exactly equal - $t_a = t_b$). Fig 2f corresponds to the atom excited in the state with a weak quantum number. Therefore, the ground state energy lies near the bottom of quantum box. Thus, its penetration over barrier becomes difficult ($t_b \approx 0$). This manifests itself in the difference between the inversion dynamics: the blue colored regions in Fig 2f show oscillations in time, however without spatial motion. The reason is in the negative inversion, which corresponds to the ground state with a vanishing tunneling probability.

Figs 5, 6 show the temporal dynamics and the spectra of the dipole current, respectively. Their qualitative behavior agreed with the corresponding inversion features. Fig 6b presents the conventional case of RO: the spectrum of the dipole current corresponds to the duplet with splitting value $2\Omega_R$ with respect to the frequency of optical transition. The central line is absent due to zero detuning (exact resonance). For RBO with $k = 0$ (Fig 6d) the independent spectrum lines are transformed into triplets with double-Bloch splitting. The next important peculiarity is the appearance of a central triplet with rather small amplitude even in the exact resonance case. For RBO with $k \neq 0$ (Figs 6c-6e) the sideband triplets transformed into multiplets with a large number of lines and Bloch frequency in the capacity of inter-line separation. For the Rydberg-atomic case (Figs d,e) the amplitude of the central triplet enhanced and has become comparable with additional lines. For the case of deep ground state (Fig c) the RO are suppressed and the central triplet became dominant.

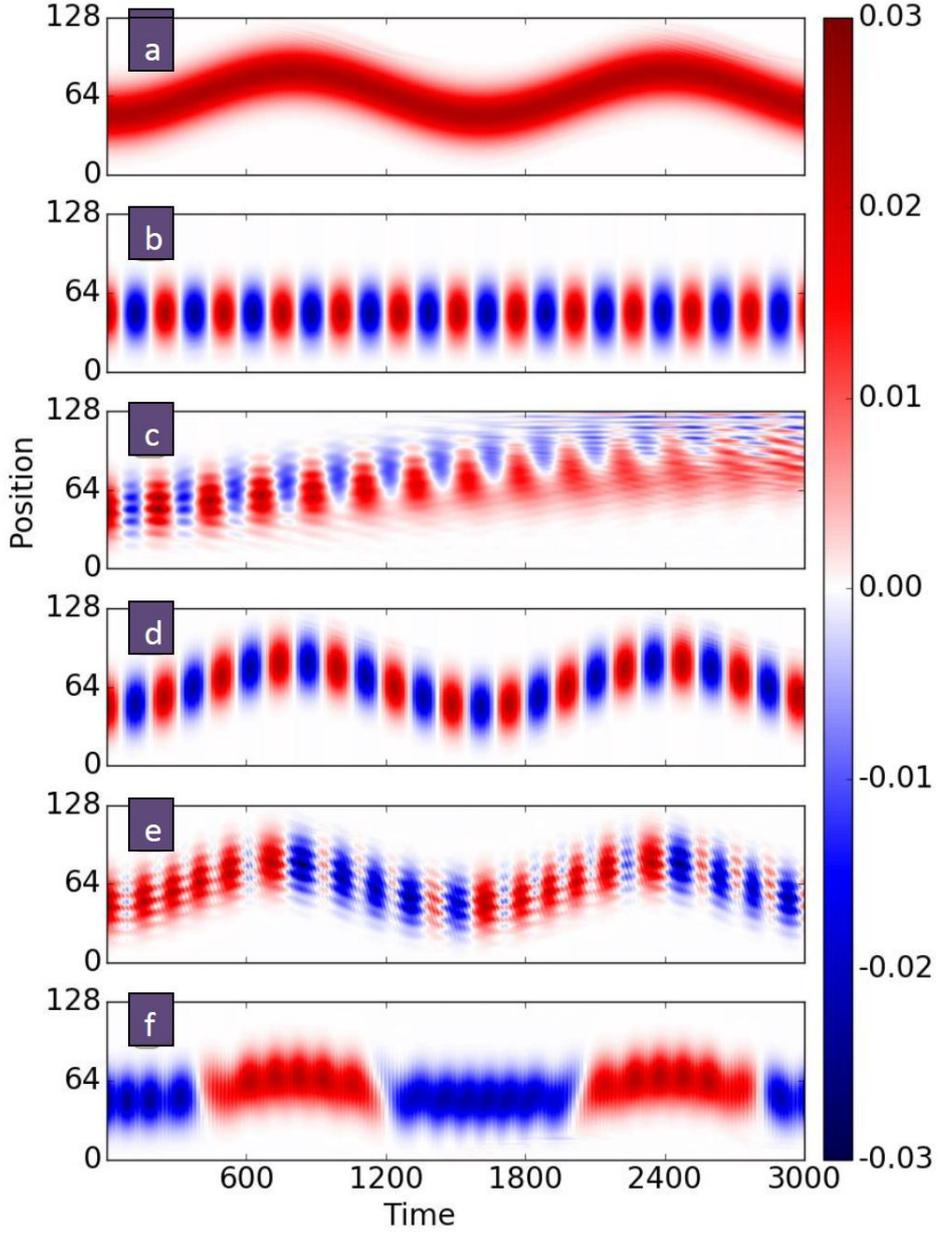

FIG.2. Space-time distribution of inversion density. a)-f) notation corresponds Table 1. Here, the quantum transition frequency is taken as the frequency unit, $\Omega_B = 3.9 \times 10^{-3}$ (corresponds $E_{Dc} = 1.95\text{kV/cm}$), $\Omega_R = 2.5 \times 10^{-2}$, $t_a = 3.5 \times 10^{-2}$, interatomic distance $a = 20nm$. The initial state of the chain is an excited single Gaussian wave packet. Gaussian initial position and width are $p' = 80$, $g = 20$, respectively, $ka = -0.624$. Number of the atoms $N = 128$.

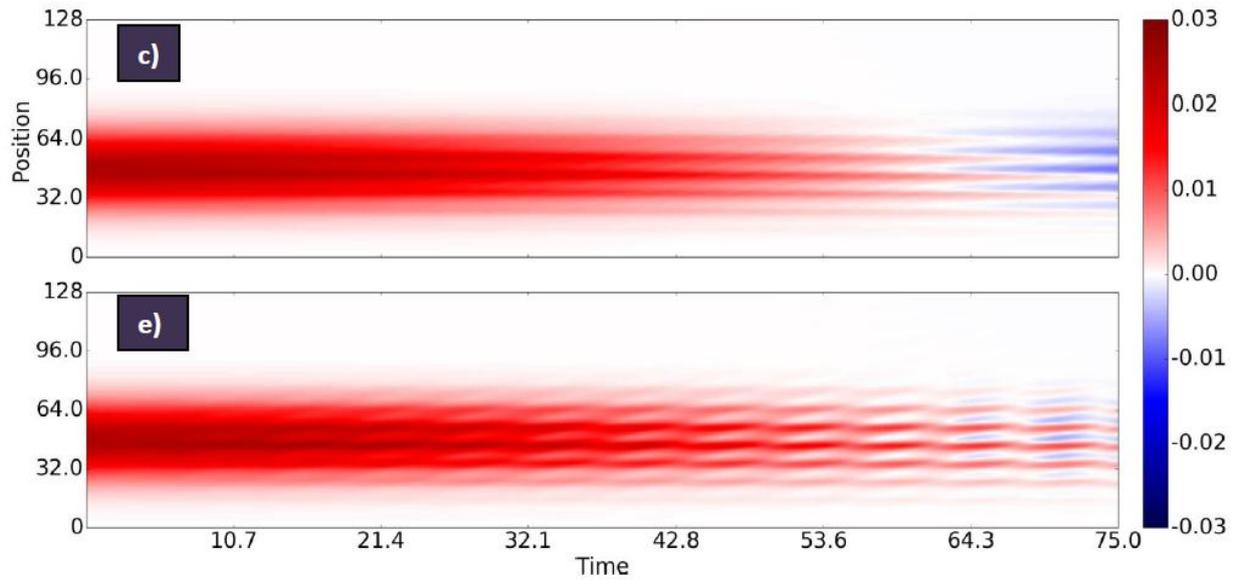

FIG. 3. Space-time distribution of inversion density for short times. c), e) notation corresponds Table 1. Spatial beatings correspond to the Floquet-harmonics in the electron diffraction by the lattice induced by the ac-field at the case of oblique incidence. As a result, the initial single Gaussian decays into the set of sub-packets, which synchronously move and coherently oscillate in the large-time regime (Fig. 2). All parameters are identical to Fig. 2

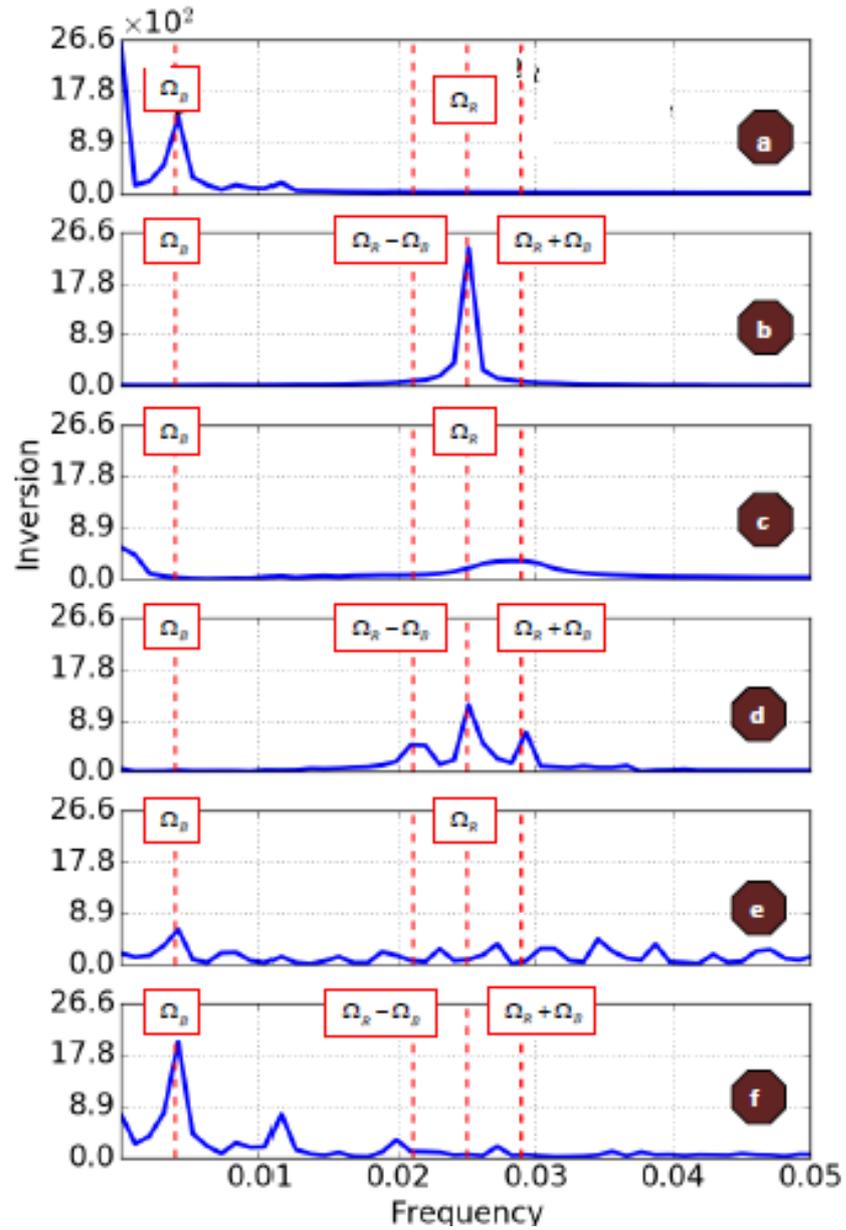

FIG.4. Frequency spectra of the inversion density. Here, a)-f) notation corresponds Table 1. The quantum transition frequency is taken as the unit. All parameters are identical to Fig. 2

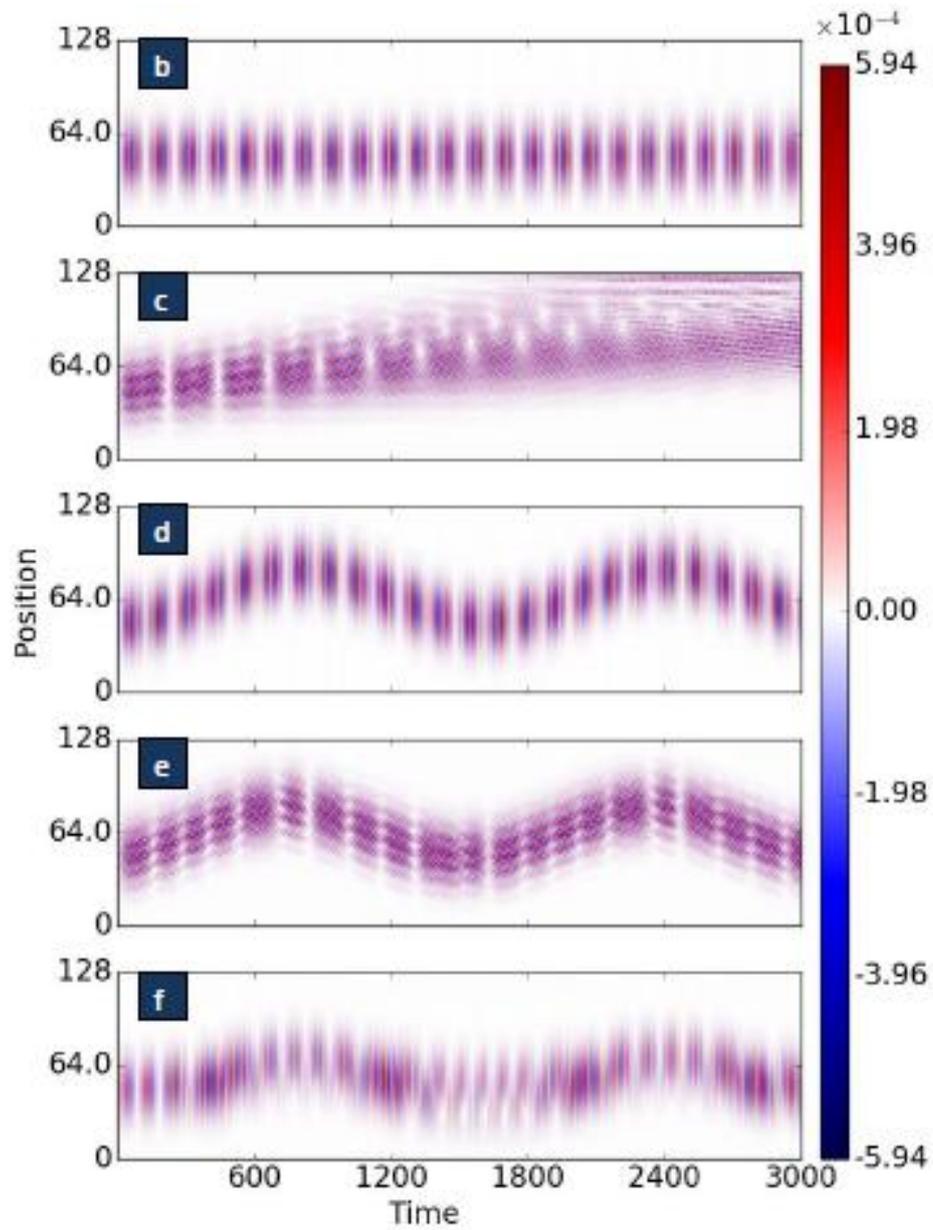

FIG.5. Space-time distribution of the density of dipole current. Here, b)-f) notation corresponds Table 1. All parameters are identical to Fig. 2

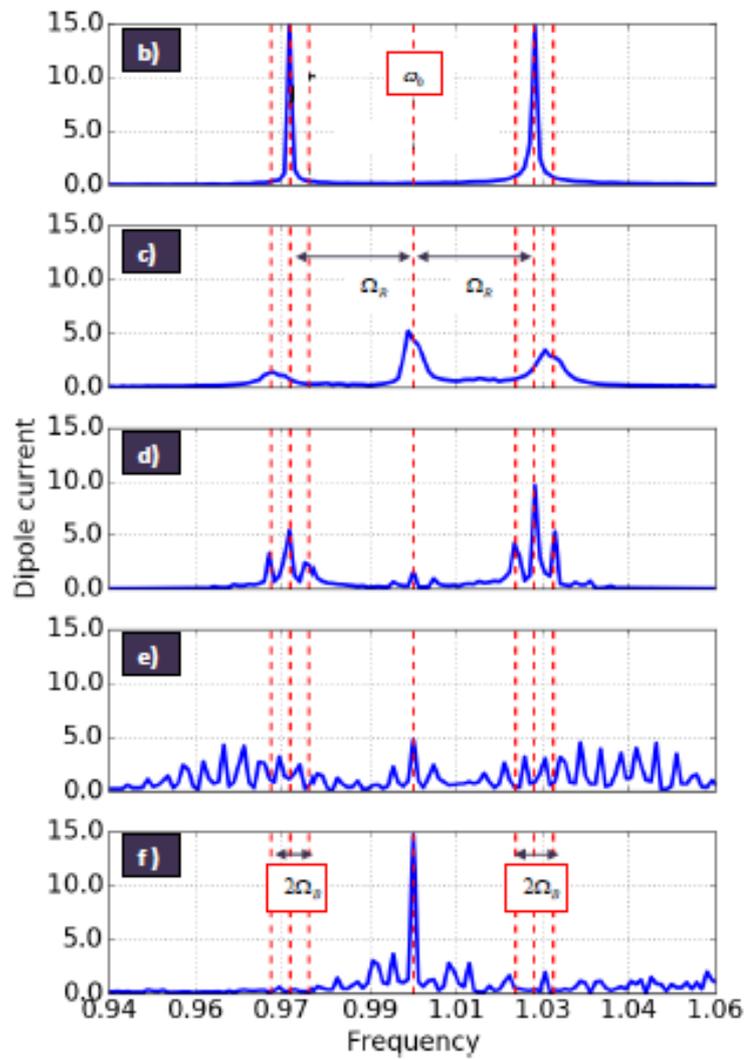

FIG.6. Frequency spectra of dipole current. Here, b)-f) notation corresponds Table 1. The quantum transition frequency is taken as the unit. All other parameters are identical to Fig. 2

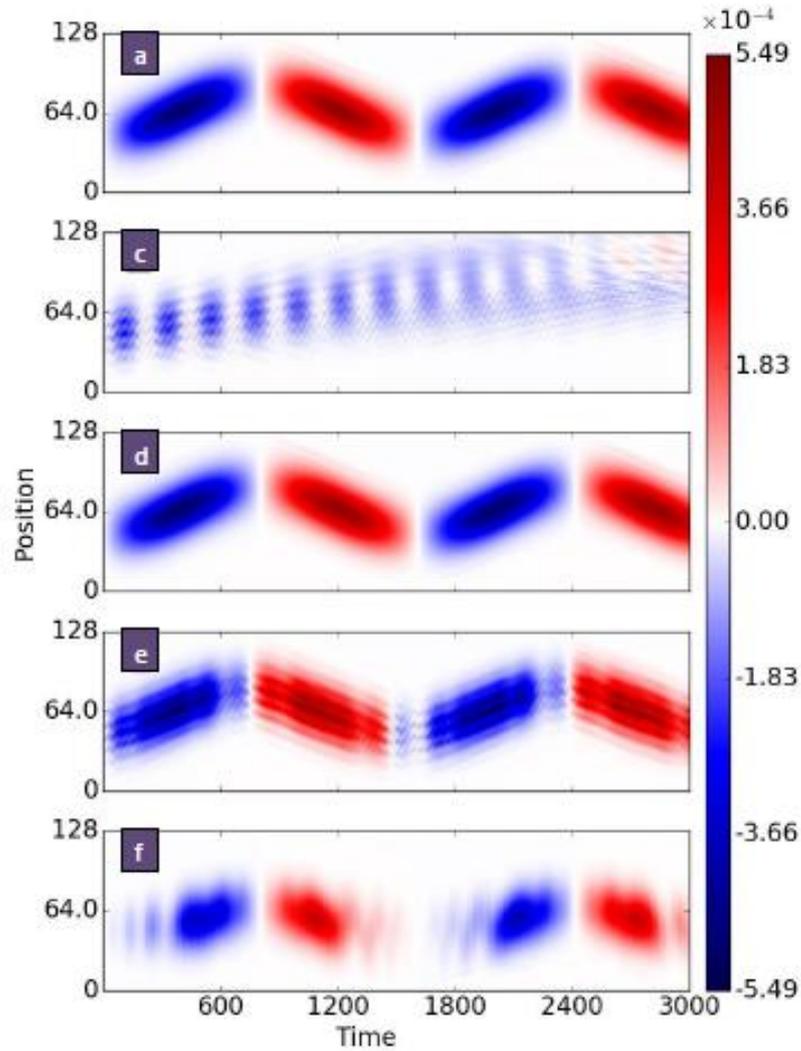

FIG.7. Space-time distribution of the tunnel current density. Here, a), c)-f) notation correspond Table 1. All parameters are identical to Fig. 2

Fig. 7 demonstrates the dynamics of tunnel current. Let us note, that it is again in agreement with inversion behavior. In particular, at Fig 7f we observe cells of zero tunnel current in the area of RBO, in contrast with Fig.7a Such cells correspond to the areas of negative inversion and stopped electron motion vat Fig 2f. The tunnel current spectrum (Fig 8) consists a dc component, the main Bloch-component and high-order Bloch lines. In all cases the second Bloch harmonic is high; its amplitude is comparable with the main line. The mutual RO and BO influence manifests itself in the enhancement of high-order harmonics in the tunnel current (Figs 8a,d-f).

Since the early days of physics, spectroscopy has been a platform for many fundamental theoretical and experimental investigations. Of course, it is impossible to give here the fundamental historical review, but short view of some important tendencies is suitable. One of the central concepts in spectroscopy is a resonance and its corresponding resonant frequency. Types of spectroscopy can also be distinguished by the nature of the interaction between the energy and the condensed matter (as examples, one may mention absorption and emission of light, Raman and Compton scattering, nuclear magnetic resonance, etc.). The first period in the making of spectroscopy is associated with atomic and molecular spectroscopy. Atoms of different elements have distinct spectra, and therefore atomic spectroscopy allows for the identification of a sample's elemental composition. The spectral lines obtained by experiment and theory can be mapped to the Mendeleev Periodic Table of the Elements. Spectroscopic studies at this time were central to the development of quantum mechanics. The combination of atoms into molecules leads to the creation of unique types of energetic states and therefore unique spectra of the transitions between these states. Molecular spectra can be obtained due by electron spin states (electron paramagnetic resonance), molecular rotations, molecular vibration and electronic states.

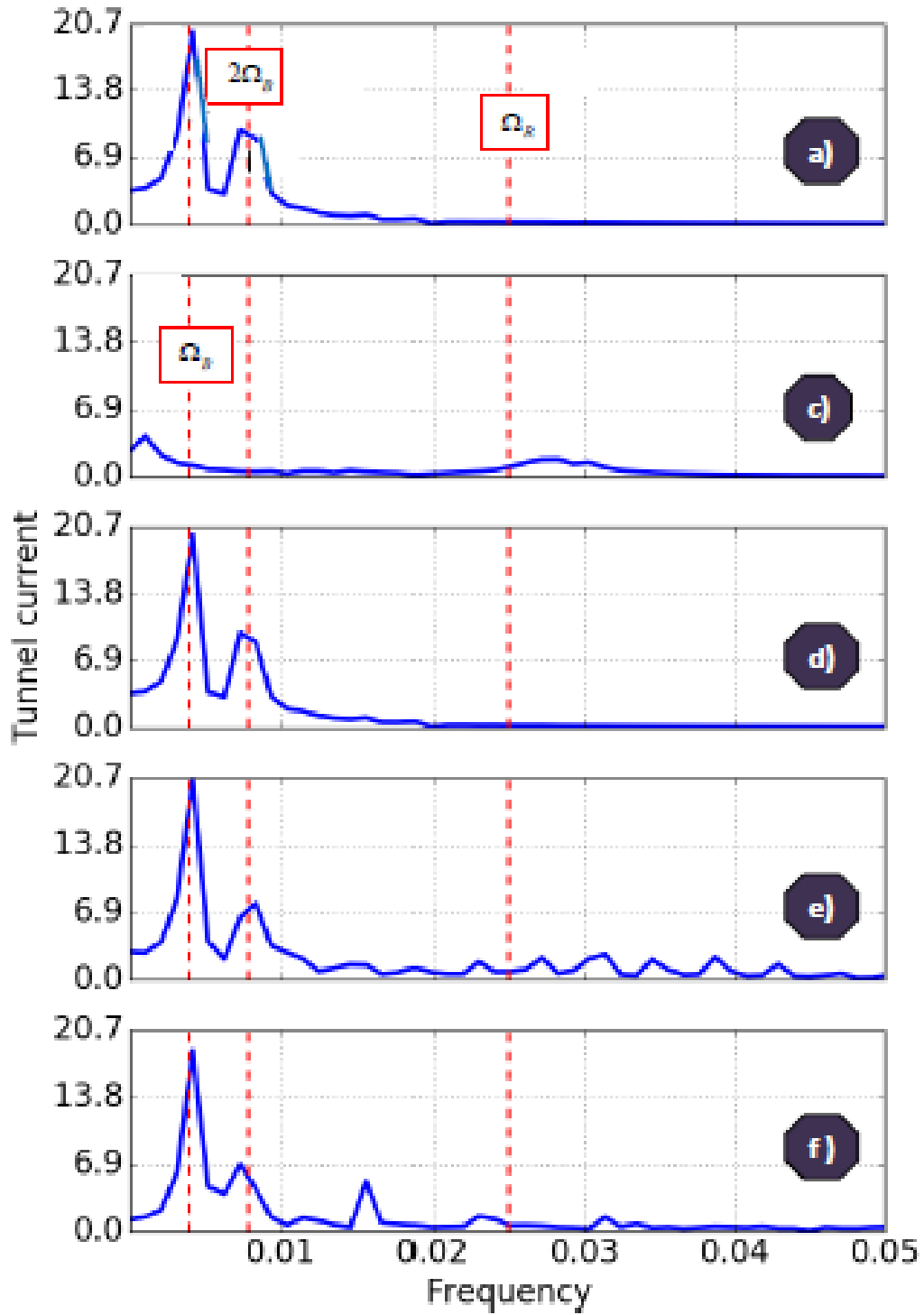

FIG.8. Frequency spectrums of tunnel current. Here, b)-f) notation corresponds Table 1. The quantum transition frequency is taken as the unit. All other parameters are identical to Fig. 2

The second period in the making of spectroscopy is associated with a wide range of applications to different types of chemical substances: gases, liquids, crystals, polymers, etc. The combination of atoms or molecules into macroscopic samples or other extended forms leads to the creation of additional energetic states. Different materials have distinct spectra. On the other hand, the dopes or lattice defects produce the characteristic spectral distortions (broadening, additional lines). Therefore, spectroscopy observations in a wide frequency range (from microwave until gamma-ray) became irreplaceable tool for the identification of a sample's structure, chemical composition, samples quality, etc in biophysics, chemistry and material engineering. Quantum mechanics provided an explanation of and theoretical framework for understanding spectroscopic observations.

Recent progress in electronics and informatics is associated with the development of nanotechnologies and synthesis of different types of nano-objects. The distinct spectra became the attribute of nano-objects of various spatial configuration, but the same chemical composition (it may be illustrated by the example of carbon nanotubes, which conductivity type (semiconductor, metal) and, correspondently, the optical spectrum depends on the nanotube radius and chiral angle). In this context, we think that the future period in the spectroscopy development whould be associated with the spectroscopy of the whole range of electronic devices or their rather large components. This trend will provide tools of the tunable spectroscopy adapted for such types of tasks, and the RBO-spectroscopy discussed above may be considered as one of the promising examples from this point of view.

## IV. ROTATION-WAVE APPROXIMATION

The physics of periodic low-dimensional lattices and their interaction with EM-field is developed with some fundamental simplifications, such as tight-binding approximation and rotating-wave approximation (RWA) [57]. In spite of the fact that these conventional models have been tested by long-term experience, their validity becomes questionable with the introduction new physical factors and corresponding degrees of freedom. Thus, their testing remains a long-standing interest. As an instructive example of such type, we can note the phenomenon of virtual photons [58], which qualitavely changes the long time dynamics of collective spontaneous emission and cannot be described by RWA. The Rabi-wave theory [22-27] is based on RWA, but adding dc-field (and BO) does not guarantee its validity for the problem at hand. Therefore, although the model developed in this paper is free of RWA, special numerical experiments with RWA will be presented in this Section and compared with direct calculations, such that the sources of potential errors can be identified.

The RWA is based on neglecting highly oscillating terms [57], and for our case equations (19), (20) may be written as follows

$$i\hbar \frac{\partial a_j}{\partial t} = \left(\delta\varepsilon - \mathbf{E}_0 \cdot e\mathbf{R}_j\right)a_j + t_a a_{j+1} + t_a^* a_{j-1} - \frac{1}{2}\omega A_0 (\mathbf{u} \cdot \mathbf{d}_{ab}) e^{i((\mathbf{k}\cdot\mathbf{e})ja - \omega t)} b_j, \qquad (28)$$

$$i\hbar \frac{\partial b_j}{\partial t} = -\left(\delta\varepsilon + \mathbf{E}_0 \cdot e\mathbf{R}_j\right)b_j + t_b b_{j+1} + t_b^* b_{j-1} - \frac{1}{2}\omega A_0 (\mathbf{u} \cdot \mathbf{d}_{ba}) e^{-i((\mathbf{k}\cdot\mathbf{e})ja - \omega t)} a_j, \qquad (29)$$

Results of the comparison of the RWA with a direct solution for inversion dynamics are shown in Figs 9-11. Two different scenarios are observed for different RBO regimes. The first one corresponds to oscillations with Rabi-frequency and smoothly various envelopes. The second one exhibits the collapse-revival picture similar to the Jaynes-Cummings dynamics [57], though the electromagnetic field is of classical origin. The reason for this is that in both cases the RO spectrum is multi-harmonic independently of its physical origin. One can see that the RWA correctly reproduces the qualitative temporal behavior of inversion (in the first or second type of the above mentioned scenarios), but allows for rather high inaccuracies in the detailed description, in contrast to the single Rabi-oscillator [57].

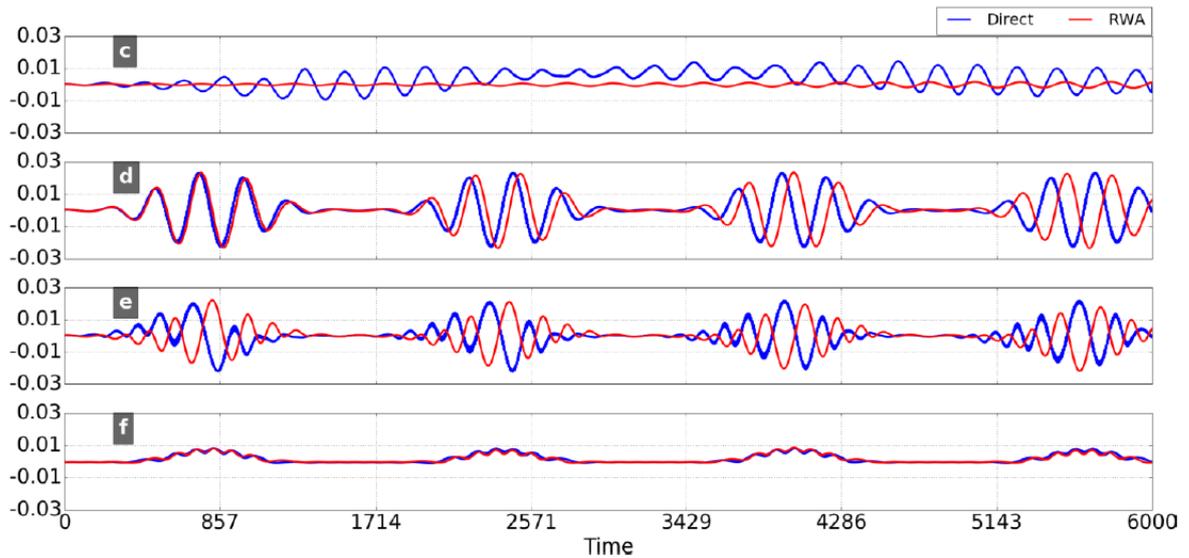

FIG. 9 Plot of inversion against a dimensionless time at the atom with number 40. . Here, c)-f) notation corresponds Table 1. Direct solution – blue line, RWA solution – red line. The initial state of the chain is an excited single Gaussian wave packet. All units and parameters are identical to Fig. 2. One can see, that RWA correctly reproduces the qualitative dynamics of inversion, but allows rather high inaccuracies in detail description.

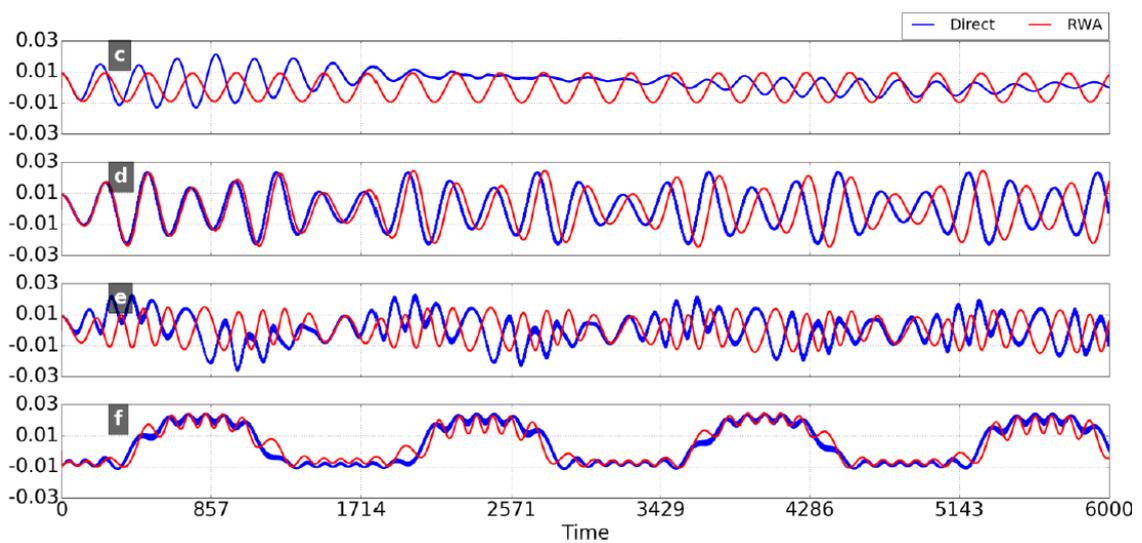

FIG. 10 Plot of inversion against a dimensionless time at the atom with number 60. Here, c)-f) notation corresponds Table 1. Direct solution – blue line, RWA solution – red line. The initial state of the chain is an excited single Gaussian wave packet. All units and parameters are identical to Fig. 2.

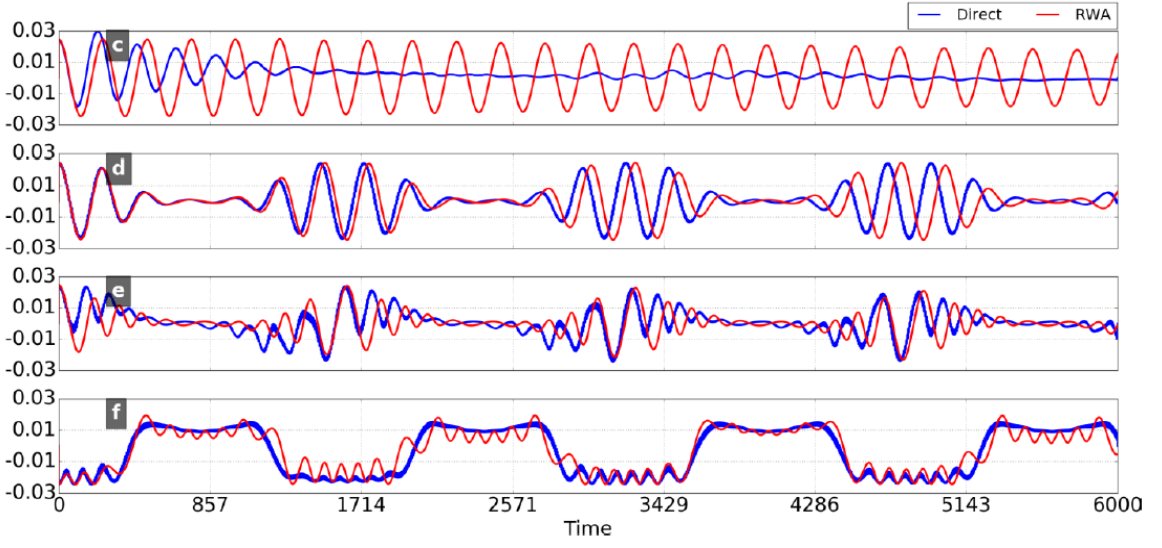

FIG. 11 Plot of inversion against a dimensionless time at the atom with number 80. Here, c)-f) notation corresponds Table 1. Direct solution – blue line, RWA solution – red line. The initial state of the chain is an excited single Gaussian wave packet. All units and parameters are identical to Fig. 2.

## V. RABI-BLOCH OSCILLATIONS AND COHERENT TRAPPING

We begin our analysis with the simplest analytical model based on the system (28,29), solving it in the limits of RWA for the case $\Omega_B = 0$. We assume for simplicity to have fulfilled an exact resonance condition $\omega = \delta\varepsilon/\hbar$. The steady state of the chain (with oscillations at this transition frequency) is

$$|\Psi(t)\rangle = \frac{1}{\sqrt{N}} \sum_j \left\{ u e^{-i\frac{\delta\varepsilon}{2\hbar}t} |a, j\rangle + v e^{i\frac{\delta\varepsilon}{2\hbar}t} |b, j\rangle \right\} e^{ihja} \quad (30)$$

where $u, v$ are unknown constant amplitudes normalized as $|u|^2 + |v|^2 = 1$, and $h$ is the unknown wavenumber. Substituting (30) to (28,29), we obtain following homogeneous system with $u, v$ as unknowns:

$$\begin{pmatrix} t_a \cos\left[\left(h + \frac{k}{2}\right)a\right] & -\frac{\Omega_R}{4} \\ -\frac{\Omega_R}{4} & t_b \cos\left[\left(h - \frac{k}{2}\right)a\right] \end{pmatrix} \begin{pmatrix} u \\ v \end{pmatrix} = 0 \quad (32)$$

Eq. (32) may be considered from different points of view: given two of three parameters $h, k, \Omega_R$, the third one is considered as a required eigen value. We assume for simplicity $k = 0$ and find the eigen-wavenumber as

$$ha = \pm \arccos\left(\frac{\Omega_R}{4\sqrt{t_a t_b}}\right) \tag{32}$$

As a result, the steady state (30) becomes

$$|\Psi(t)\rangle = \frac{1}{\sqrt{N}\sqrt{1+\frac{t_b}{t_a}}} \sum_j \left\{ \sqrt{\frac{t_b}{t_a}} e^{-i\frac{\delta\varepsilon}{2\hbar}t} |a,j\rangle + e^{i\frac{\delta\varepsilon}{2\hbar}t} |b,j\rangle \right\} e^{ihja} \tag{33}$$

This state exists in the finite area of the ac-field values satisfied the inequality $|\Omega_R| \leq 4\sqrt{t_a t_b}$. It may be considered as an analog of the inversion trapped state in three-level atom $\Lambda$-configuration [57]. This analogy is non-complete; because of the inversion trapping in our case is not perfect. The reason is non-vanishing probability amplitude for excited state $|a,j\rangle$. But this value may be done arbitrary small via the choice of relation between tunneling frequencies ($t_b \ll t_a$). Thus, in contrast with three-level atom, the chain of two-level atoms in the state (33) is not perfectly transparent to the incident field, but the reachable values of the photon absorption are arbitrary small if the tunneling in the ground state is suppressed enough as compared with the excited one.

Let us now analyze a wave packet analog of the coherent trapped state (33). This is done based on the RWA-free eqs. (28,29), and solved numerically with initial conditions

$$\begin{pmatrix} a_j(0) \\ b_j(0) \end{pmatrix} = g \begin{pmatrix} \sqrt{t_b/t_a} \\ 1 \end{pmatrix} e^{-\frac{(j-j')^2 a^2}{\sigma^2}} e^{ihja} \tag{34}$$

for different relations between the system parameters including dc field appearance ($g$ is normalizing coefficient). Numerical results are shown at Fig. 12. The case i) corresponds to the ordinary BO. No real oscillations are observed because of their weak amplitude which is due to the small value of the tunnel penetration $t_b$. Case ii) corresponds to the RO. Here, no oscillations are observed because of the initial state trapping. The monotonic movement is stipulated by the non-zero initial momentum of the wave-packet $h$. Case iii) corresponds to RBO (involving simultaneous action of dc and ac fields). One can observe Rabi-like oscillations between ground and excited states. In contrast with the ordinary RO scenario, their frequency is equal to the Bloch-frequency and defined by the value of the dc field. BO are clearly visible too due to the tunneling through the exited level with high penetration value, in spite of the small contribution of this state in the total wave function. Case iv) corresponds to the Rabi-wave, where the picture has changed dramatically due to the trapping of the initial state. The trapped state exhibits spatial oscillations of the wavefunction that are due to diffraction by the grating and are induced via the field travelling over the array axis ($k \neq 0$), however propagates without inter-level transitions. The case v) demonstrates both diffraction and inter-level transitions with the Bloch-frequency being the result of the simultaneous action of dc and ac fields with $k \neq 0$. In conclusion, it is found that the dc field for the two-level atomic chain is able to break down the coherence of the trapped states. As a result, the ground and excited states become coupled, which leads to the inter-level oscillations accompanied by Bloch-oscillations. This is a novel physical mechanism, which holds promise for a number of potential applications in nano-electronics, such as electrically controlled nano-antennas and networks.

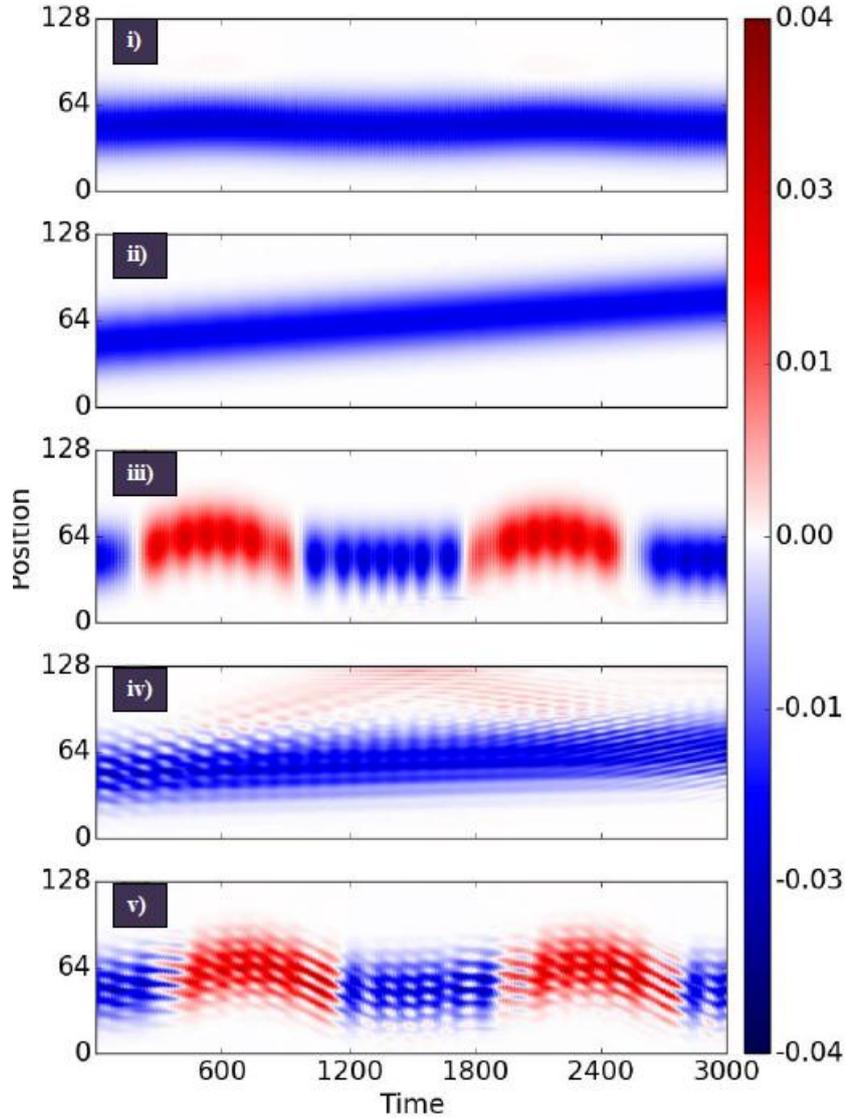

FIG.12. Space-time distribution of inversion density for the initial state in the form of the Gaussian analog of the inversion trapped state given by Eq. (34). Here, the quantum transition frequency is taken as the frequency unit. Gaussian initial position and width are $p' = 80$, $g = 20$, respectively, Here, interatomic distance $a = 20 nm$, $t_a = 3.5 \times 10^{-2}$ eV, $t_b = 3.5 \times 10^{-3}$ eV number of the atoms is 128. i) BO-case: $\Omega_B = 3.9 \times 10^{-3}$, $\Omega_R = 0$. No BO have been observed, because of their small amplitude due to the weak tunneling penetration over barrier for the ground state; ii) RO-case: $\Omega_B = 0$, $\Omega_R = 2.5 \times 10^{-2}$, standing wave ($k = 0$). No RO have been observed, because of the special interference mechanism of inversion trapping. Progressive motion of the wave-packet is dictated by the non-zero value of its initial momentum. iii) RBO-case: $\Omega_B = 3.9 \times 10^{-3}$, $\Omega_R = 2.5 \times 10^{-2}$, standing wave ($k = 0$). Both, RO and BO are observed and progressive motion is suppressed due to the combined action of ac- and dc-fields; iv) RO-case: $\Omega_R = 2.5 \times 10^{-2}$, $\Omega_B = 0$, travelling wave ($ka = -0.624$). No RO have been observed, similar to the case ii). Progressive motion of the wave-packet is accompanied the spatial beatings dictated by diffraction on the lattice induced by the travelling wave; v) RBO-case: $\Omega_B = 3.9 \times 10^{-3}$, $\Omega_R = 2.5 \times 10^{-2}$, travelling wave ($ka = -0.624$). Both, RO and BO are observed similar to the case iii), and accompanied by the spatial beatings similar to the case iv);

## VI. SUMMARY AND OUTLOOK

By way of summary and open questions, the control of RO is a problem of long-standing interest. As seen above, it is a platform for rich physics, and not in vain, it was raised in Ref [63] the question: "What Rabi oscillations can teach us?" The same question may be addressed BO. This question may be answered in due to course when the synthesis into RBO inveils new and excited capabilities. We hope, that this paper has provided the initial steps towards this goal.

In conclusion, we have developed a model of RBO driven by the superposition of dc and ac fields in a one-dimensional chain of two-level atoms coupled via tunneling. We have presented a derivation of the equations of motion for the dc field, which is homogeneous along the chain, and plane-wave ac field obliquely incident with an arbitrary direction. These equations enable a strong atom-field coupling modeling based on the Wannier basis. Examples of its solution have been obtained by purely numerical means, demonstrating the spatial-temporal densities of inversion, with dipole current and tunnel current calculated for the system, initiated with a Gaussian packet state.

.    As a central result, we have shown, that the system dynamics is characterized by a superposition of the oscillating inter-atomic intra-band transitions via electron tunneling (BO) and the intra-atomic inter-band quantum transitions (RO). In spite of such components belonging to different areas of frequency, the RBO-dynamics is characterized by strong mutual interaction of Bloch and Rabi-components, which qualitatively changes the physical picture of both of them. In particular, we have seen that inversion exhibits the collapse-revival behavior in the inversion in the classical ac field. In this case the BO-propagation plays the role of the photonic dress in the Jaynes-Cummings model [57].

Identifying such qualitative behavior makes the validity of such conventional approximation as RWA subject of scrutiny. In this regard, the used a model that is free of RWA, hovever calculations using the assumption of the RWA have been done too and compared with the exact numerical solution. As a result, the RWA was overall shown to reproduce the numerical results qualitatively to a fairly high degree of accuracy through both the approximate solution and a direct numerical implementation of the theory.

Although we have focused on the spectral properties of RBO, the mutual influence of RO and BO interactions was considered in detail. It was shown, that the RBO-spectrum consists of low-frequency component (THz) governed by the tunneling, and high-frequency (optical) component dictated by the inter-level transitions. The complicated behavior of spatial-temporal dynamics leads to appearance of a lot of additional spectral lines whose amplitudes are compared with the main one.

This opens the door for exciting possibilities of applications in a new type of spectroscopy in nano-electronics and electrical control in nano-devices.

## ACKNOWLEDGMENTS

G.Ya. acknowledges support from the project FP7-PEOPLE-2013-IRSES-612285 CANTOR.

## APPENDIX: DERIVATION OF FORMULA FOR TUNNELING CURRENT

For calculating the tunneling current we introduce the operator of the particle number in the $j$-th atom $\hat{N}_j = |a_j\rangle\langle a_j| + |b_j\rangle\langle b_j|$ and formulate the equation of continuity

$$\hat{J}_j - \hat{J}_{j-1} = -e\frac{d\hat{N}_j}{dt} \tag{A.1}$$

where $\hat{J}_j$ is the current density operator in the *j*-th atom and the left-hand part in (A.1) is a discrete analog of Div in 1D-case. Using Heisenberg equation for the operator $\hat{N}_j$, we rewrite (A.1) in the form

$$\hat{J}_j - \hat{J}_{j-1} = -i\frac{e}{\hbar}\left[\hat{H}, \hat{N}_j\right] \tag{A.2}$$

The tunneling currents over excited and background energy levels are independent, thus $\hat{J}_{\text{Tunneling},j} = \hat{J}^{(a)}_{\text{Tunneling},j} + \hat{J}^{(b)}_{\text{Tunneling},j}$. Using (A.2), we obtain

$$\hat{J}^{(a,b)}_{\text{Tunneling},j} - \hat{J}^{(a,b)}_{\text{Tunneling},j-1} = -i\frac{e}{\hbar}\left[\hat{H}^{(a,b)}_{\text{Tunneling}}, \hat{N}_j\right] \tag{A.3}$$

where

$$\hat{H}^{(a)}_{\text{Tunneling}} = \sum_j (t_a |a,j\rangle\langle a,j+1| + t_a^* |a,j\rangle\langle a,j-1|) \tag{A.4}$$

is a component of Hamiltonian (10), chargeable for the tunneling at the excited level (the similar equation may be written for the Hamiltonian $\hat{H}^{(b)}_{\text{Tunneling}}$). Using (A.4), we calculate the commutator in the right-hand part of (A.3) and obtain

$$\hat{J}^{(a)}_{\text{Tunneling},j} - \hat{J}^{(a)}_{\text{Tunneling},j-1} = -iet_a \left(|a_{j-1}\rangle + |a_{j+1}\rangle\right)\langle a_j| + H.c. \tag{A.5}$$

It corresponds to the operator of the tunneling current at the excited level

$$\hat{J}^{(a)}_{\text{Tunneling},j} = -iet_a |a_j\rangle\langle a_{j+1}| + H.c. \tag{A.6}$$

The observable value of the tunneling current is

$$J^{(a)}_{\text{Tunneling},j}(t) = \left\langle \hat{J}^{(a)}_{\text{Tunneling},j} \right\rangle = -iet_a a_j^*(t) a_{j-1}(t) + c.c. \tag{A.7}$$

Using the approximation $a_{j-1}(t) - a_j(t) \approx \left(a_{j+1}(t) - a_{j-1}(t)\right)/2$ and adding the similar support of the ground level, we obtain the relation (22). This relation corresponds to the well-known definition of the probability flow $\mathbf{j} = i\hbar\left(\Psi \text{grad}\Psi^* - c.c.\right)/2m$ in 3D continuous case [60].